\documentclass [12pt] {article}
\usepackage{graphicx}
\usepackage{epsfig}
\begin {document}

 \begin{center}

 {\Large {\bf TETRAQUARKS AND PENTAQUARKS IN STRING MODELS}} \\

 \vskip 1.5 truecm

 {\bf Yu. M. Shabelski and M. G. Ryskin}\\

 \vskip 0.5 truecm

 Petersburg Nuclear Physics Institute, Gatchina, St.Petersburg,

 Russia \\

 \vskip 1. truecm

 E-mail: shabelsk@thd.pnpi.spb.ru\\
 E-mail: ryskin@thd.pnpi.spb.ru

 \end{center}

 \vskip 1.5 truecm

 \begin{center}

 {\bf ABSTRACT}

 We consider the production and decay of multiquark systems in the
 framework of string models where the hadron structure is determined
 by valence quarks together with string junctions. We show that the
 low mass multiquark resonances can be very narrow.

 \end{center}

 \vskip 1cm

 PACS. 25.75.Dw Particle and resonance production

 \vskip 2.5 truecm

 \newpage

 \section{Structure of multiquark systems}

 In the string models baryons are considered as configurations consisting
 of three strings (related to three valence quarks) connected at the point
 called "string junction" (SJ) \cite{Artru}-\cite{Khar}. The string
 junction has a nonperturbative origin in QCD. Many phenomenological
 results (some of them will be discussed below) were obtained in this
 approach 25 years ago \cite{IOT}, \cite{RV}, \cite{IOT3}-\cite{Noda}.

 In QCD hadrons are composite bound state configurations built up
 from the quark $\psi_i(x), i = 1,...N_c$ and gluon $G^{\mu}_a(x),
 a = 1,...,N_c^2-1$ fields. In the string models the meson wave
 function has the form of "open string" \cite{Artru,RV}, as it is
 shown in Fig.~1a.

 \begin{figure}[htb]
 \centering
 \includegraphics[width=.5\hsize]{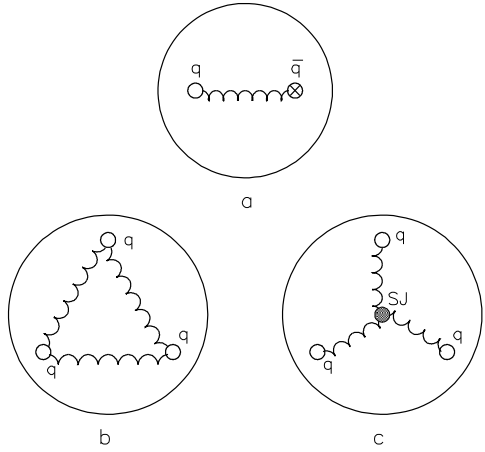}
 \caption{Composite structure of a meson (a) and a baryon (b) and (c)
 in string models. Quarks are shown by open points and antiquarks by
 crossed points.}
 \end{figure}

 The meson wave function (here and below we present only its colour
 structure) reads as

 \begin{equation}
 M = \bar{\psi}^i(x_1) \Phi_i^{i'}(x_1,x_2)\psi_{i'}(x_2) \;.
 \end{equation}

 \begin{equation}
 \Phi_i^{i'}(x_1,x_2) = \left[ T\exp \left(g \int_{P(x_1,x_2)}
 A_{\mu}(z) dz^{\mu}\right) \right]_i^{i'} \;,
 \end{equation}

 where the field $A_{\mu}$ is the matrix in colour space :

 \begin{equation}
 A_{\mu} = \sum_a t_a A^a_{\mu} \;.
 \end{equation}

 In the last equation $P(x_1,x_2)$ represents a path from $x_1$ to
 $x_2$ which looks like an open string with ends at $x_1$ and $x_2$.

 For the baryons there exist two possibilities, "triangle", or
 $\Delta$ connection shown in Fig.~1b and "star", or $Y$ connection
 shown in Fig.~1c. The last variant is considered as the most
 interesting. Here a baryon is considered as configurations consisting
 of three strings attached to three valence quarks and connected in a
 point called the "string junction" (SJ) \cite{Artru,RV}.
 Such a picture is confirmed by lattice calculations \cite{latt}.
 The correspondent wave function can be written as

 \begin{equation}
 B = \psi_i(x_1) \psi_j(x_2)\psi_k(x_3) J^{ijk} \;,
 \end{equation}

 \begin{equation}
 J^{ijk} = \Phi^i_{i'}(x_1,x)\Phi_{j'}^j(x_2,x)\Phi^k_{k'}(x_3,x)
 \epsilon^{i'j'k'} \;,
 \end{equation}

 Such baryon wave function can be defined as a "star" or "Y" shape
 and it is preferable \cite{Artru,RV} in comparison with "triangle"
 ("ring") or "$\Delta$" shape\footnote{Strictly speaking we can not
 build up the "$\Delta$" configuration with the help of a string 
Eq.~(2). In this ("$\Delta$") case the colour flux produced by the quark 
is divided between two strings. That is we need the fractional colour
 factor (like $g/2$) in the power of the exponent in Eq.~(2).}.

 The wave function of an antibaryon has the form

 \begin{equation}
 \bar{B} = \psi^i(x_1) \psi^j(x_2)\psi^k(x_3) J_{ijk} \;.
 \end{equation}

 The operators $J^{ijk}$ and $J_{ijk}$ differ by the position of
 colour indices that gives possibility of annihilation of
 $B\bar{B}$ pair into mesons.

 The presented picture leads to several phenomenological predictions
 \cite{ait}-\cite{SJ1}.
 In particular, there exist the rooms for exotic states, such as
 glueboll, or gluonium ("closed string"), Fig.~2a, \cite{RV,Ven}.

 \begin{equation}
 Glueboll = Tr \left[ T\exp \left(g \int_{P(closed)}
 A_{\mu}(z) dz^{\mu}\right) \right] \;.
 \end{equation}

 \begin{figure}[htb]
 \centering
 \vskip -1cm
 \includegraphics[width=.5\hsize]{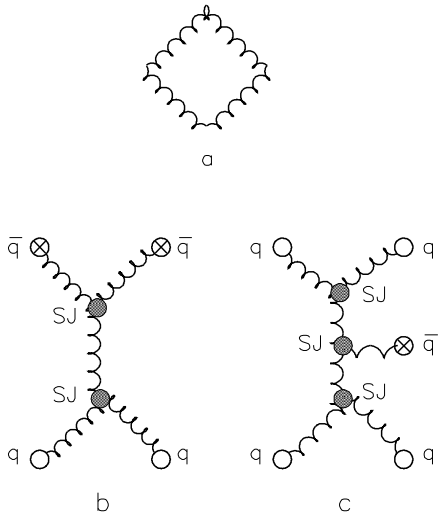}
 \caption{Exotic states: glueboll (a), 4-quark meson (tetraquark)
 $M_4 = qq\bar{q}\bar{q}$ (b) and 5-quark baryon (pentaquark)
 $B_5 = qqqq\bar{q}$ (c) in string models. Quarks are shown by open
 points and antiquark by crossed points.}
 \end{figure}

 The multiquark bound states, such as 4-quark meson, Fig.~2b, pentaquark,
 Fig.~2c, etc. also can exist. Without specified model it is impossible
 to say about the sign of the correspondent binding energy, i.e. are
 they the bound states or not. However we can expect that the part of a
 particle momentum carried out by gluons in the case of multiquark states
 should be larger than for usual particles, Figs.~1a, 1c due to the
 larger number of string junctions.

 Similarly to Eqs.~(4), (5), the meson $M_4$ (tetraquark) wave function
 can be written as

 \begin{equation}
 M_4 = \psi_i(x_1) \psi_j(x_2) J^{ijm} \times
 \bar{\psi}^k(x_3) \bar{\psi}^l(x_4) J_{kln} \times
 \Phi_m^n(x_0^{(1)},x_0^{(2)}) \;,
 \end{equation}

 and the pentaquark wave function

 \begin{equation}
 B_5 = \psi_i(x_1) \psi_j(x_2) J^{ijn} \times
 \psi_l(x_4) \psi_m(x_5) J^{lmq} \times
 \bar{\psi}^k(x_3) \Phi_k^{k'}(x_3,x_0^{(2)}) \epsilon_{nqk'} \;,
 \end{equation}

 Specified form of colour wave functions of $M_4$ and $B_5$
 presented above should results in weak mixing of them with
 $qq\bar{qq}$ and $qqqq\bar{q}$ states produced radiatively from
 usual mesons and baryons.

 \section{Tetraquark and pentaquark production}

 The probability of SJ pair production in small space-time region is
 rather small. In the case of pentaquark production an estimation can be
 taken from the ratio of $\pi^-p \to \bar{p}d$ reaction cross section to
 the total inelastic one. Experimentally \cite{Flam} it is of the order
 of $10^{-3} - 10^{-4}$ at $\sqrt{s} = 2.9 - 3.2$ GeV. Note that these
 energies are rather close to the $\bar p d$ threshold. The momenta of
 the final nucleons are about 200 - 300 MeV. So the expected
 suppression ($\sim 0.01$) due to the low probability to form
 the deuteron is not too strong and it looks reasonable to assume that
 an additional factor of ($0.01\ -\ 0.1)$ is caused by the SJ pair
 production.\\

 One way to avoid the smallness is to use the
 initially prepared SJ. So we can produce a tetraquark in
 $\bar{p}p$ collisions and a pentaquark in $\bar{p}d$ collisions at
 rather small energy. In the first case we can expect the annihilation
 of one $q\bar{q}$ pair that corresponds to the planar Dual
 Topological Unitarization diagram which is the leading in $1/N_c$
 expansion. In the last case we expect the annihilation of two
 $q\bar{q}$ pairs and the additional smallness about $10^{-2}$ which
 is connected with probability to find a proton and a neutron in the
 deuteron at small distances.

 In the case of, say, pentaquark production in $\pi^-p$ collisions
 the ratio of signal to background should be worse in comparison with
 $\bar{p}d$ interactions.

 \section{Tetraquark and pentaquark decay}

 Let us start from the case when the mass of a meson $M_4$ (tetraquark),
 Fig.~2b, is large enough. In the considered the simplest mode of a meson
 $M_4$ decay is the breaking of the string between the points $x_0^{(1)}$
 and $x_0^{(2)}$ with production of a light $q\bar{q}$ pair that should
 result in decay of $M_4$ into $B\bar{B}$ state. The decay into two
 mesons should be lesser preferable, that is supported also by the
 triality analysis \cite{IOT3}. Similarly a baryon $B_5$ (pentaquark),
 Fig.~2c, with high enough mass should decay preferably into
 $BB\bar{B}$ state via breaking of two strings and production of two
 $q\bar{q}$ pairs\footnote{The probability of two SJ annihilation
(or production) is assumed to be small\cite{IOT3,IOT5,CN2}. It was
argue in Ref.~\cite{CN1} that the smallness of SJ annihilation may be
caused by the small size ($\sim 0.2-0.4$ fm) of the ``junction''.}.

The situation becomes more complicate in the case of low mass
 multiquark states \cite{DPP,Nar}. Due to the completeness condition we can
 consider only real hadrons in the intermediate states. So we can imagine
 that now tetraquark decays firstly into virtual $B\bar{B}$ pair with
 their subsequent annihilation into mesons. Similarly, in the case of
 pentaquark decay the intermediate $BB\bar{B}$ system will results in
 the system of a baryon together with several mesons. The important point
 is the suppression factor coming from every virtual baryon with
 4-momentum $k$. In comparison with the normal width of $N^*$ resonances
 ($\Gamma\sim 150\ -\ 200$ MeV) here we have the suppression due to the
 loop which provides the annihilation of the virtual $B\bar B$ pair into
 mesons. Besides the numerically small factor $1/4\pi$ the loop contains
 the baryons far off the mass-shell. This leads to the suppression of
 the order of $(2m_\pi)^2/(m_B^2-k^2)\sim 1/10$. That is we expect the
 supression of the decay amplitude $\sim 1/100$, i.e. we expect the
 width of about $10^{-4}\cdot 200$ MeV = 20 KeV or even smaller, since
 the $B\bar B$ pair mainly annihilates into a rather high
 multiplicity pion states and there should be an additional suppression
 due a limited phase space avaliable for the final pions. Small width of 
pentaquarks was also obtained \cite{IofO,Og} in the framework of QCD sum 
rules approach.

 There exist the possibility to search tetraquarks and pentaquarks
 in the decay modes with strange particles, for example
 $K^0_s + \Lambda + X$, where $s\bar{s}$ pair can be produced in the
 process of string breaking. The constituent mass of a strange quark is not much larger
 than that of a light ($u,d$) quarks.
  The correspondent decay mode should be
 suppressed by the factor $10^{-1}$, but the background should be
 suppressed more significantly.

 Moreover, since the events with the $K^0_s$ or $\Lambda$ decay
 are easy to select experimentally, the decay modes with strange
 particles or/and the reactions of a strange pentaquark  (the one which
 contains strange quark) production look as an attractive way to search
 for such a multiquark states.\\

 The small width of the pentaquark (tetraquark) can explain the
 problems of their search, see, e.g. reviews \cite{Dan,Bur}.
 Indeed, it is hard to find a very narrow state performing the usual
 partial wave analysis. On the other hand the cross section of the
 inclusive pentaquark (or tetraquark) production is expected to be very
 small in the processes where there is no three SJ in the initial state
 and one needs to create two new SJ.

 Therefore, it looks more perspective to search for the pentaquark in
 the $\bar p d$ annihilation.

 We conclude that the existance of tetraquarks and pentaquarks is
 natural in the string models. The production probability can be enhanced
 in special combinations of a beam and a target. The light multiquark
 systems can have rather large mean life time (in the scale of strong
 interactions) that can produce an additional problems for their
 experimental search.

 We are grateful to A. K. Likhoded, L. N. Lipatov and V. Yu. Petrov for
 useful discussions. This paper was supported by DFG grant GZ: 436 RUS
 113/771/1-2 and, in part, by grants RSGSS-1124.2003.2 and
 PDD (CP) PST.CLG980287.

 \end{document}